\documentclass{IEEEtran}
\usepackage{cite}
\usepackage{amsmath,amssymb,amsfonts}
\usepackage{algorithmic}
\usepackage{graphicx}
\usepackage{textcomp}
\usepackage{color}
\def\BibTeX{{\rm B\kern-.05em{\sc i\kern-.025em b}\kern-.08em
T\kern-.1667em\lower.7ex\hbox{E}\kern-.125emX}}    

\begin{document}

\title{A Survey of Approximate Quantile Computation on Large-scale Data (Technical Report)}

\author{
Zhiwei Chen$^{1}$,
Aoqian Zhang$^{2}$\\
$^{1}${School of Software, Tsinghua University, Beijing, 100084, China}\\
$^{2}${Cheriton School of Computer Science, University of Waterloo, Waterloo, Ontario, Canada N2L 3G1}
}

\maketitle

\begin{abstract}
As data volume grows extensively, data profiling helps to extract metadata of large-scale data. However, one 
kind of metadata, order statistics, is difficult to be computed because they are not mergeable or incremental. 
Thus, the limitation of time and memory space does not support their computation on large-scale data. In this 
paper, we focus on an order statistic, quantiles, and present a comprehensive analysis of studies on 
approximate quantile computation. Both deterministic algorithms and randomized algorithms that compute 
approximate quantiles over streaming models or distributed models are covered. Then, multiple techniques for 
improving the efficiency and performance of approximate quantile algorithms in various scenarios, such as 
skewed data and high-speed data streams, are presented. Finally, we conclude with coverage of existing packages 
in different languages and with a brief discussion of the future direction in this area.
\end{abstract}

%----------------------------------------------------
\section{Introduction}\label{sect:introduction}
Data profiling is a set of activities to describe the metadata about given data \cite{DBLP:journals/vldb/AbedjanGN15,DBLP:journals/vldb/Song0Y13}. 
It is crucial for data analysis, especially for large-scale data. It helps researchers to 
understand data distribution \cite{DBLP:journals/tcs/HeCCW15}, discover duplicates \cite{DBLP:journals/isci/SongZ014,DBLP:journals/tkdd/WangSCYC17}, 
detect anomalies \cite{DBLP:journals/pvldb/ZhangS0Y17}, determine thresholds \cite{DBLP:journals/tkde/Song0C14,DBLP:journals/tods/Song011}, etc. 
Such information provides guidance for other data preprocessing work such as data cleaning \cite{DBLP:conf/sigmod/SongZWY15,DBLP:conf/sigmod/ZhangSW16}, 
which can subsequently improve the performance of data mining dramatically \cite{DBLP:conf/kdd/SongLZ15}. 
When preprocessing large-scale data, 
data profiling is attached great importance to and faces its own challenges. Because of large data size, 
classic brutal methods are not applicable any more for their intolerable complexity of both time and space. Researchers 
have spent decades on figuring out new ways to compute the metadata which can be calculated easily on small data. 
The metadata can be divided into two categories based on scalability: aggregation statistics and order statistics \cite{DBLP:journals/siamcomp/GuhaM09}.

Aggregation statistics are named for their property that they are mergeable and incremental, which makes them relatively easy to be computed no matter how large the data is. For examples, sum, mean values, standard deviations, min or max values are all aggregation statistics. For streaming models \cite{DBLP:conf/pods/BabcockBDMW02, DBLP:journals/fttcs/Muthukrishnan05}, where data elements come one by one with time, we can trace and update aggregated results covering all arrived data by incrementing new results continuously. Time complexity and space complexity are both $O(1)$. As for distributed models \cite{DBLP:journals/tit/Segall83}, where data are stored in nodes of a distributed network, the overall aggregation statistics can be obtained by merging results from each node. The total communication cost of this computation is $O(|v|)$, where $|v|$ is the number of network nodes. However, order statistics, such as quantiles, heavy hitters, etc., do not preserve such property. So, we cannot compute them by merging existing results with newly produced results in a straight way. In order to compute them, many customized data structures or storage structures are proposed for these order statistics, trying to turn them into a mergeable or incremental form in some way.

In this summary, we focus on one order statistic, quantiles. They help to generate the description of the data distributions without parameters. 
In other words, they are able to reflect the cumulative distribution function (cdf), thus the probability distribution function (pdf), of data at low 
computational cost. Pdf is widely used in data cleaning and data querying. For example, in data cleaning, it is applied to demonstrate the distance distribution among values of the same attribute so as to identify misplaced attribute values \cite{sun2020swapping}. And in data querying, it helps to set an appropriate correlation filter, improving efficiency for set correlation query over set records in databases \cite{DBLP:journals/jcst/GaoSCW16}. Therefore, quantiles are regarded as one of the most fundamental and most important statistics in data quality analysis in both theory and practice. 
For instance, many data analysis tools, including Excel, MATLAB, Python, etc., have quantile-computing functions as 
built-in components or libraries. In the Sawzall language, which is the basic for all Google’s log data analysis, 
quantile is one of the seven basic statistic operators defined, along with sum, max, top-k, etc. \cite{DBLP:journals/sp/PikeDGQ05}. Besides, 
quantiles are widely used in data collection and running-state monitoring in sensor networks \cite{DBLP:conf/sigmod/JohnsonCKMSS04, DBLP:conf/sensys/ShrivastavaBAS04}. 
When a dataset contains dirty values, compared with mean values and standard deviations, quantiles and median absolute deviations are more objective and more accurate to reflect data center and data deviation \cite{leys2013detecting}. They are less sensitive to outliers. In temporal data, where imprecise timestamps are prevalent, even if some timestamps are delayed very long or have inconsistent granularity, quantiles are still able to specify appropriate temporal constraints on time interval, helping to clean the data \cite{DBLP:journals/pvldb/SongC016}. In addition, quantile algorithms have been widely used as subroutines to resolve more complicated problems, such as equi-depth histograms \cite{DBLP:books/sp/16/GreenwaldK16} and dynamic geometric computations \cite{DBLP:conf/stoc/Indyk04}.

A quantile is the element at a certain rank in the dataset after sort. Algorithmic studies can be traced back to 1973 at least when linear-time selection was invented \cite{DBLP:journals/jcss/BlumFPRT73}. In classic methods of computing $\phi$-quantile over a dataset of size $N$, where $\phi \in (0, 1)$, first we sort all elements and then return the one ranking $\lfloor\phi{N}\rfloor$. Its time complexity is $O(N\log{N})$ and space complexity is $O(N)$ obviously. However, in large-scale data, the method is infeasible under restrictions of memory size. 
Munro \emph{et al.} has proved that any exact quantile algorithm with $p$-pass scan over data requires at least $\Omega(N^{1/p})$ space \cite{DBLP:journals/tcs/MunroP80}. Besides, in streaming models, 
e.g., over streaming events \cite{DBLP:journals/tkde/0001SZLS16}, 
quantile algorithms should also be streaming.
That is, they are permitted to scan each element only once and need to update quantile answers instantaneously when receiving new elements. There is no way to compute quantiles exactly under such condition. Thus, approximation is introduced in quantile computation. Approximate computation is an efficient way to analyze large-scale data under restricted resources \cite{DBLP:journals/csur/Mittal16b}. On one hand, it raises computational efficiency and lower computational space. On the other hand, large scale of the dataset can dilute approximation effects. Large-scale data is usually dirty, which also makes approximate quantile endurable and applicable in industry. Significantly, the scale of data is relative, based on the availability of time and space. So, the rule about how to choose between exact quantiles and approximate quantiles differs in heterogeneous scenarios, depending on the requirement for accuracy and the contradiction between the scale of data and that of resources. When the cost of computing exact quantiles is intolerable and the results are not required to be totally precise, approximate quantiles are a promising alternative.

\begin{figure}[htbp]
    \includegraphics[width=0.55\columnwidth]{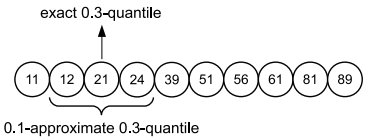}
    \caption{An example of a $\epsilon$-approximate $\phi$-quantile, where $\epsilon=0.1$ and $\phi=0.3$.}
    \label{fig:quantile-example}
\end{figure}

We denote approximation error by $\epsilon$. A $\epsilon$-approximate $\phi$-quantile is any element whose rank is between $r-\epsilon{N}$ and 
$r+\epsilon{N}$ after sort, where $r = \lfloor\phi{N}\rfloor$. For example, we want to calculate 
$0.1$-approximate $0.3$-quantile of the dataset ${11, 21, 24, 61, 81, 39, 89, 56, 12, 51}$. 
As shown in Figure~\ref{fig:quantile-example}, we sort the elements as ${11, 12, 21, 24, 39, 51, 56, 61, 81, 89}$ 
and compute the range of the quantile's rank, which is $[(0.3 - 0.1) \times 10, (0.3 + 0.1) \times 10] = [2, 4]$. 
Thus the answer can be one of ${12, 21, 24}$. In order to further reduce computation space, approximate quantile 
computation is often combined with randomized sampling, making the deterministic computation becomes randomized. 
In such case, another parameter $\delta$, or randomization degree, is introduced, meaning the algorithm answers 
a correct quantile with a probability of at least $1 - \delta$.

There are 3 basic metrics to assess an approximate quantile algorithm \cite{DBLP:journals/vldb/LuoWYC16}:
\begin{itemize}
  \item \textbf{Space complexity} It is necessary for streaming algorithms. Due to the limitation of memory, only algorithms using sublinear space are applicable \cite{DBLP:conf/escape/GangulyM07}. It corresponds to communication cost in distributed models. In a distributed sensor network, communication overhead consumes more power and limits the battery life of power-constrained devices, such as wireless sensor nodes \cite{DBLP:conf/vldb/DeshpandeGMHH04}. So, quantile algorithms are aimed at low space complexity or low communication cost.
  \item \textbf{Update time} It is the time spent on updating quantile answers when new element arrives. Fast updates can improve the user experience, so many streaming algorithms take update time as a main consideration. 
  \item \textbf{Accuracy} It measures the distance between approximate quantiles and ground truth. Intuitively, the more accurate an algorithm is, the more space and the longer time it will consume. They are on the trade-off relationship. We use approximation error, maximum actual approximation error and average actual approximation error as quantitative indicators to measure accuracy.
\end{itemize}

We collected and studied researches about approximate quantile computation, then completed this survey. 
The survey includes various algorithms, varying from data sampling to data structure 
transformation, and techniques for optimization. Important algorithms are listed in Table~\ref{tab:algorithm}. 
The remaining parts of this survey are organized as follows: In Section~\ref{sect:deterministic algorithms}, 
we introduce deterministic quantile algorithms over both streaming models and distributed models. 
Section~\ref{sect:randomized algorithms} discusses randomized algorithms \cite{DBLP:journals/csur/MotwaniR96}. Section~\ref{sect:improvement} 
introduces some techniques and algorithms for improving the performance and efficiency of quantile algorithms 
in various scenarios. Section~\ref{sect:tools} presents a few off-the-shelf tools in industry for quantile 
computation. Finally, Section~\ref{sect:conclusions} makes the conclusion and proposes interesting 
directions for future research. 

\begin{table*}[htbp]
  \caption{Approximate quantile algorithms}
  \label{tab:algorithm}
  \setlength{\tabcolsep}{3pt}
  \begin{center}
  \begin{tabular}{|p{100pt}|p{100pt}|p{100pt}|p{150pt}|}
    \hline
    Algorithm & Randomization & Model & Space complexity / Communication cost\\
    \hline
    MRL98 \cite{DBLP:conf/sigmod/RajagopalanML98} & Deterministic & Streaming & $O(\frac{1}{\epsilon}\log^2(\epsilon{N}))$ \\
    GK01 \cite{DBLP:conf/sigmod/GreenwaldK01} & Deterministic & Streaming & $O(\frac{1}{\epsilon}\log(\epsilon{N}))$ \\
    Lin SW \cite{DBLP:conf/icde/LinLXY04} & Deterministic & Streaming & $O(\frac{1}{\epsilon}\log(\epsilon^2N)+\frac{1}{\epsilon^2})$ \\
    Lin n-of-N \cite{DBLP:conf/icde/LinLXY04} & Deterministic & Streaming & $O(\frac{1}{\epsilon^2}\log^2(\epsilon{N}))$ \\
    Arasu FSSW \cite{DBLP:conf/pods/ArasuM04} & Deterministic & Streaming & $O(\frac{1}{\epsilon}\log\frac{1}{\epsilon}\log{N})$ \\
    Arasu VSSW \cite{DBLP:conf/pods/ArasuM04} & Deterministic & Streaming & $O(\frac{1}{\epsilon}\log\frac{1}{\epsilon}\log(\epsilon{N})\log{N})$ \\
    GK-UN \cite{DBLP:journals/access/LiangLL19} & Deterministic & Streaming & $O(\frac{1}{\epsilon}\log(\epsilon{PC(S^u)}))$ \\
    Q-digest \cite{DBLP:conf/sensys/ShrivastavaBAS04} & Deterministic & Distributed & $O(\frac{1}{\epsilon}|v|\log\sigma)$ \\
    GK04 \cite{DBLP:conf/pods/GreenwaldK04} & Deterministic & Distributed & $O(\frac{|v|}{\epsilon}\log^2{N})$ \\
    Cormode05 \cite{DBLP:conf/sigmod/CormodeGMR05} & Deterministic & Distributed & $O(\frac{|v|}{\epsilon^2}\log{N})$ \\
    Yi13 \cite{DBLP:journals/algorithmica/YiZ13} & Deterministic & Distributed & $O(\frac{|v|}{\epsilon}\log{N})$ \\
    MRL99 \cite{DBLP:conf/sigmod/MankuRL99} & Randomized & Streaming & $O(\frac{1}{\epsilon}\log^2\frac{1}{\epsilon})$ \\
    Agarwal13 \cite{DBLP:journals/tods/AgarwalCHPWY13} & Randomized & Streaming & $O(\frac{1}{\epsilon}\log^{1.5}\frac{1}{\epsilon})$ \\
    Felber15 \cite{DBLP:conf/approx/FelberO15} & Randomized & Streaming & $O(\frac{1}{\epsilon}\log\frac{1}{\epsilon})$ \\
    Karnin16 \cite{DBLP:conf/focs/KarninLL16} & Randomized & Streaming & $O(\frac{1}{\epsilon}\log\log\frac{1}{\epsilon\delta})$ \\
    Huang11 \cite{DBLP:conf/sigmod/HuangWYL11} & Randomized & Distributed & $O(\frac{1}{\epsilon}\sqrt{|v|h})$ \\
    Haeupler18 \cite{DBLP:conf/podc/HaeuplerMS18} & Randomized & Distributed & $O(\log\log{N}+\log\frac{1}{\epsilon})$ \\
    \hline
  \end{tabular}
  \end{center}
\end{table*}

\section{Deterministic Algorithms}
\label{sect:deterministic algorithms}
An algorithm is deterministic while it returns a fixed answer given the same dataset and query condition. 
Furthermore, quantile algorithms are classified based on their application scenarios. 
In streaming models, where data elements arrive one by one in a streaming way, algorithms are required 
to answer quantile queries with only one-pass scan, given the data size $N$ \cite{DBLP:conf/sigmod/RajagopalanML98} or 
not \cite{DBLP:conf/vldb/AlsabtiRS97, DBLP:conf/kdd/ChenLP00, DBLP:conf/sigmod/GreenwaldK01, DBLP:conf/pods/ArasuM04, DBLP:journals/access/LiangLL19}. Except to answering quantile queries for
all arrived data, Lin \emph{et al.} \cite{DBLP:conf/icde/LinLXY04} concentrates on tracing quantiles for the most recent $N$ elements 
over a data stream. In distributed models, where data or statistics are stored in distributed architectures such 
as sensor networks, algorithms are proposed to merge quantile results from child nodes using as low communication 
cost as possible to reduce energy consumption and prolong equipment life \cite{DBLP:conf/sensys/ShrivastavaBAS04, DBLP:conf/pods/GreenwaldK04, 
DBLP:conf/sigmod/HuangWYL11, DBLP:journals/tods/AgarwalCHPWY13, DBLP:journals/algorithmica/YiZ13}.

\subsection{Streaming Model}
\label{sect:deterministic-streaming}
The most prominent feature of streaming algorithms is that all data are required to be scanned only once. Besides, the length of the data stream 
may be uncertain and can even grow arbitrarily large. Thus, classic quantile algorithms are infeasible for streaming models because of the limitation 
of memory. Therefore, the priority of approximate quantile algorithms for streaming models is to minimize space complexities. In 2010, Hung \emph{et al.} \cite{DBLP:conf/faw/HungT10} have proved that any comparison-based $\epsilon$-approximate quantile algorithm over streaming models needs space complexity of at least $\Omega(\frac{1}{\epsilon}\log\frac{1}{\epsilon})$, which sets a lower bound for these algorithms. 

Both Jain \emph{et al.} \cite{DBLP:journals/cacm/JainC85} and Agrawal \emph{et al.} \cite{DBLP:conf/comad/AgrawalS95} proposed algorithms to compute quantiles with one-pass scan. However, neither of them clarified the upper or lower bound of approximation error. In 1997, Alsabti \emph{et al.} \cite{DBLP:conf/vldb/AlsabtiRS97} improved the algorithm and came up with 
a version with guaranteed error bound, referred to as ARS97. Its basic idea is sampling and it includes the following steps:
\begin{enumerate}
  \item Divide the dataset into $r$ partitions.
  \item For each partition, sample $s$ elements and store them in a sorted way.
  \item Combine $r$ partitions of data, generating one sequence for querying quantiles.
\end{enumerate}
ARS97 is targeted at disk-resident data, rather than streaming models. Nevertheless, its idea to partition the entire dataset and maintain a 
sampled sorted sequence for quantile querying inspires quantile algorithms over streaming models afterwards.

\begin{figure}[htbp]
    \includegraphics[width=\columnwidth]{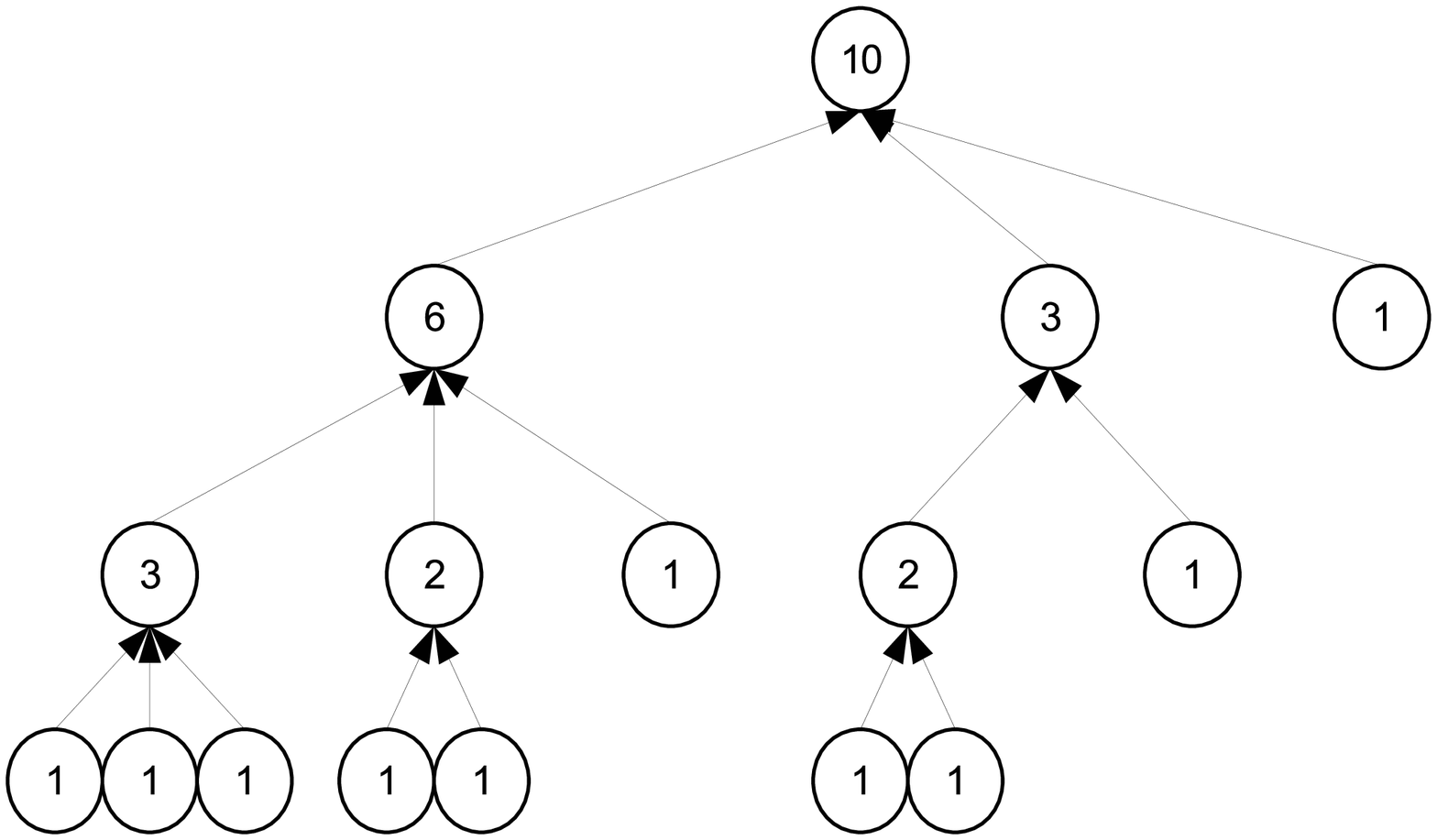}
    \caption{The collapsing strategy with 3 buffers. A node corresponds to a buffer and its number denotes the weight.}
    \label{fig:MRL98-schema}
\end{figure}

The inspired algorithm is referred to as MRL98 proposed by Manku \emph{et al.} \cite{DBLP:conf/sigmod/RajagopalanML98}. It requires the prior knowledge of the 
length $N$ of data stream. Similar with ARS97, MRL98 divides the data stream into $b$ blocks, samples $k$ elements from each block and puts 
them into $b$ buffers. Each buffer $X$ is given a weight $w(X)$, representing the number of elements covered by this 
buffer. The algorithm consists of 3 operations:
\begin{itemize}
  \item \textbf{NEW} Put the first $bk$ elements into buffers successively and set their weights to $1$.
  \item \textbf{COLLAPSE} Compress elements from multiple buffers into one buffer. Specifically, each element from an input buffer $X_i$ 
  would be duplicated $w(X_i)$ times. Then these duplicated elements are sorted and merged into a sequence, where $k$ elements are selected at 
  regular intervals and stored in the output buffer $Y$, whose weight $w(Y) = \sum_i{w(X_i)}$
  \item \textbf{OUTPUT} Select an element as the quantile answer from $b$ buffers.
\end{itemize}
NEW and OUTPUT are straightforward, so the algorithm's space complexity depends mainly on how to trigger COLLAPSE. 
MRL98 proposed a tree-structure trigger strategy. Each buffer $X$ is assigned a height $l(X)$ and $l$ is set to $min_i{l(X_i)}$. 
$l(X)$ is set as the following standard:
\begin{itemize}
  \item If only one buffer is empty, its height is set to $l$.
  \item If there are two or more empty buffers, their heights are set to $0$.
  \item Otherwise, buffers of height $l$ are collapsed, generating a buffer of height $l + 1$.
\end{itemize}
By tuning $b$ and $k$, MLR98 can narrow the approximation error within $\epsilon$. Figure~\ref{fig:MRL98-schema} 
demonstrates the trigger strategy when $b = 3$. The height of the strategy tree is logarithmic, thus the space complexity 
is $O(\frac{1}{\epsilon}\log^2(\epsilon{N}))$.

The limitation of MRL98 is that it needs to know the length of the data stream at first. However, in more cases, the 
length is uncertain and may even grow arbitrarily large. In such case, Greenwald \emph{et al.} \cite{DBLP:conf/sigmod/GreenwaldK01} 
proposes celebrated GK01 algorithm, distinguished by its innovative data structure, $Summaries$, or $S$ for short. The basic idea is 
that when $N$ increases, the set of $\epsilon$-approximate answers for querying $\phi$-quantile expands as well, so 
correctness can be retained even if removing some elements. $S$ is a collection of tuples in the form of  
$(v_i, g_i, \Delta_i)$, where $v_i$ is the element with the property $v_i \leq v_{i+1}$, $g_i$ and $\Delta_i$ are $2$ 
integers satisfying following conditions:
\begin{equation}
  \sum_{j \leq i}g_j \leq r(v_i) + 1 \leq \sum_{j \leq i}g_j + \Delta_i \label{eq:2-1}
\end{equation}
\begin{equation}
  g_i + \Delta_i \leq \lfloor 2\epsilon{N} \rfloor \label{eq:2-2}
\end{equation}
$r(v_i)$, whose bounds are guaranteed by (\ref{eq:2-1}), is the ground truth of $v_i$'s rank. Obviously, there are at 
most $g_i + \Delta_i - 1$ elements between $v_{i-1}$ and $v_i$. (\ref{eq:2-2}) makes sure that the range is within 
$\lfloor 2\epsilon{N} \rfloor - 1$. Therefore, for any $\phi \in (0, 1)$, there always exists a tuple $(v_i, g_i, 
\Delta_i)$, where $r(v_i) \in [\lfloor(\phi-\epsilon)N\rfloor, \lfloor(\phi+\epsilon)N\rfloor]$, and thus $v_i$ is a 
$\epsilon$-approximation $\phi$-quantile. To find the approximate quantile, we can find the least $i$ satisfying:
\begin{equation}
  \sum_{j \leq i}g_j + \Delta_i > 1 + \lfloor\epsilon{N}\rfloor + max_i(g_i + \Delta_i) / 2 \label{eq:2-3}
\end{equation}
and return $v_{i-1}$ as the answer. $S$ also supports several operations: 
\begin{itemize}
  \item \textbf{INSERT} Search $S$ to find the least $i$ so that $v_i > v$ and insert $(v, 1, \lfloor 2\epsilon{N} \rfloor)$ right before 
  the tuple $t_i=(v_i, g_i, \Delta_i)$.
  \item \textbf{DELETE} Update $g_{i+1}$ as $g_{i+1} = g_{i+1} + g_i$ and delete $t_i$. To maintain (\ref{eq:2-2}), $t_i$ is removable only 
  when it satisfies
  \begin{equation}
    g_i + g_{i+1} + \Delta_{i+1} \leq \lfloor 2\epsilon{N} \rfloor \label{eq:2-4}
  \end{equation}
  \item \textbf{COMPRESS} It can be found that DELETE is adding $g$ of the deleting tuple to that of its predecessor. So we can delete 
  multiple successive tuples, $t_{i+1}, t_{i+2},..., t_{i+k}$ at the same time by updating $g_i$ as 
  $g_i = g_i + g_{i+1} + g_{i+2} + \cdots + g_{i+k}$ and removing them. GK01 proposed a complicated COMPRESS strategy to reduce the size 
  of $S$ as small as possible: it executes when $t_i, t_{i+1},...,t_{i+k}$ are removable on the arrival of every $\frac{1}{2\epsilon}$ elements.
\end{itemize}
It proves that the maximum size of $S$ is $\frac{11}{2\epsilon}\log(2\epsilon{N})$. So, its space complexity is 
$O(\frac{1}{\epsilon}\log(\epsilon{N}))$. Additionally, the summary structure is also applicable in a wide range 
such as Kolmogorov-Smirnov statistical tests \cite{DBLP:conf/bigdataconf/Lall15} and balanced parallel computations \cite{DBLP:conf/sigmod/TaoLX13}. 

GK01 is for computing quantiles over all arrived data. Sometimes quantiles of the most recent $N$ elements in a stream are 
required. Lin \emph{et al.} \cite{DBLP:conf/icde/LinLXY04} expanded GK01 and proposed two algorithms for such case: SW model 
and n-of-N model.

\begin{figure}[htbp]
    \includegraphics[width=\columnwidth]{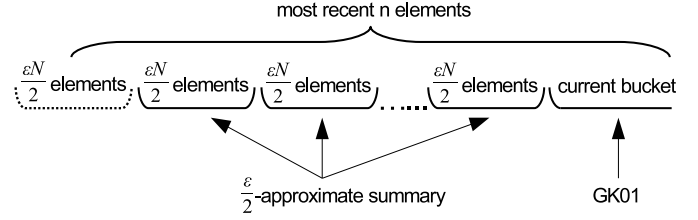}
    \caption{Structure of SW model}
    \label{fig:Lin-SW}
\end{figure}

SW model is for answering quantiles over the most recent $N$ elements instantaneously, where $N$ is predefined. 
The model puts the most recent $N$ elements into several buckets in their arriving order. Rather than original elements, 
each buckets stores a $Summary$ \cite{DBLP:conf/sigmod/GreenwaldK01} covering $\frac{\epsilon{N}}{2}$ successive elements. 
The buckets have 3 states as illustrated in Figure~\ref{fig:Lin-SW}:
\begin{itemize}
  \item A bucket is active when its coverage is less than $\frac{\epsilon{N}}{2}$. At this time, it maintains a 
  $\frac{\epsilon}{4}$-approximate $S$ computed by GK01.
  \item A bucket is compressed when its coverage reaches $\frac{\epsilon{N}}{2}$. The $\frac{\epsilon}{4}$-approximate $S$ would be compressed 
  to $\frac{\epsilon}{2}$-approximate $S$ by an algorithm COMPRESS.
  \item A bucket is expired if it is the oldest bucket when the coverage of all buckets exceeds $N$. Once expired, the bucket is removed 
  from the bucket list.
\end{itemize}

By maintaining and merging buckets, SW model answers quantile queries of elements covered by all unexpired buckets. However, when an old bucket 
is just expired and the active bucket has not been full, the coverage of all unexpired buckets $N'$ is less than $N$. The difference between $N$ 
and $N'$ is at most $\lfloor\frac{\epsilon{N}}{2}\rfloor-1$. 
An algorithm LIFT was proposed to resolve the problem: if $0 \leq N-N' \leq \lfloor\frac{\epsilon{N}}{2}\rfloor$, it can convert a 
$\frac{\epsilon}{2}$-approximate $S'$ covering $N'$ elements to a $\epsilon$-approximate $S$ covering $N$ elements. Its worst case space 
complexity is $O(\frac{1}{\epsilon}\log(\epsilon^2N)+\frac{1}{\epsilon^2})$.

\begin{figure}[htbp]
    \includegraphics[width=\columnwidth]{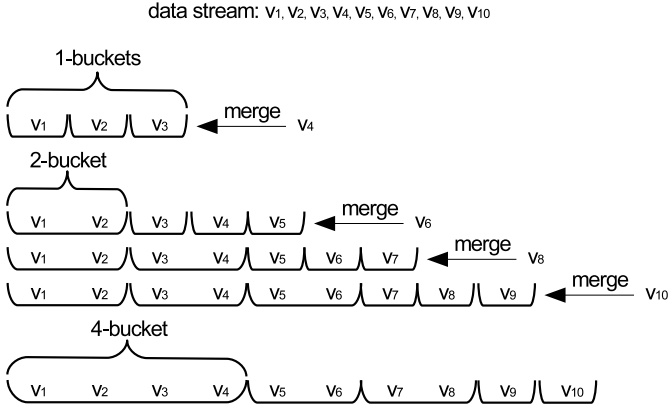}
    \caption{Structure of the EH-partition technique in n-of-N model, where $\lambda = 0.5$.}
    \label{fig:EH-bucket}
\end{figure}

The distinction of n-of-N model is that it answers quantile queries instantaneously over the most recent $n$ elements where $n$ is any integer 
not larger than a predefined $N$. It takes advantage of the EH-partition technique \cite{DBLP:journals/siamcomp/DatarGIM02} as shown in 
Figure~\ref{fig:EH-bucket}. The technique classifies buckets, marking buckets at level $i$ as $i-bucket$ whose $S$ covers all elements 
arriving since the bucket's timestamp. There are at most $\lceil\frac{1}{\lambda}\rceil+1$ buckets at each level, where $\lambda \in (0, 1)$. 
When a new element arrives, the model creates a $1-bucket$ and sets its timestamp to the current timestamp. When the number of $i-bucket$s 
reaches $\lceil\frac{1}{\lambda}\rceil+2$, 
the two oldest buckets at level $i$ are merged into a $2i-bucket$ carrying the oldest timestamp iteratively until buckets at all levels 
are less than $\lceil\frac{1}{\lambda}\rceil + 2$. Because each bucket covers elements arriving since its timestamp, merging two buckets is equal 
to removing the later one. Lin \emph{et al.} also proved that for any bucket $b$, its coverage $N_b$ satisfies
\begin{equation}
  N_b-1 \leq \lambda{N} \label{eq:2-5}
\end{equation}
In order to guarantee that quantiles are $\epsilon$-approximate, $\lambda$ is set to $\frac{\epsilon}{\epsilon+2}$. Each bucket preserves a 
$\frac{\epsilon}{2}$-approximate $S$. Quantiles are queried as follows:
\begin{enumerate}
  \item Scan the bucket list until finding the first bucket $b$ making $N_b \leq n$.
  \item Use LIFT to convert $S_b$ to a $\epsilon$-approximate $S$ covering $n$ elements. 
  \item Search $S$ to find the quantile answer.
\end{enumerate}
According to (\ref{eq:2-5}), $n - N_b \leq \frac{\epsilon{N_b}}{\epsilon+2}$. Mark its predecessor bucket as $b'$ and we have $N_{b'} > n$, 
thus $n - N_b \leq N_{b'} - N_b - 1$, and furthermore $n - N_b \leq \lfloor\frac{\epsilon{n}}{2}\rfloor$. So, LIFT can be applied 
to $S_b$. The worst case space complexity is $O(\frac{1}{\epsilon^2}\log^2(\epsilon{N}))$. 

\begin{figure}[htbp]
    \includegraphics[width=\columnwidth]{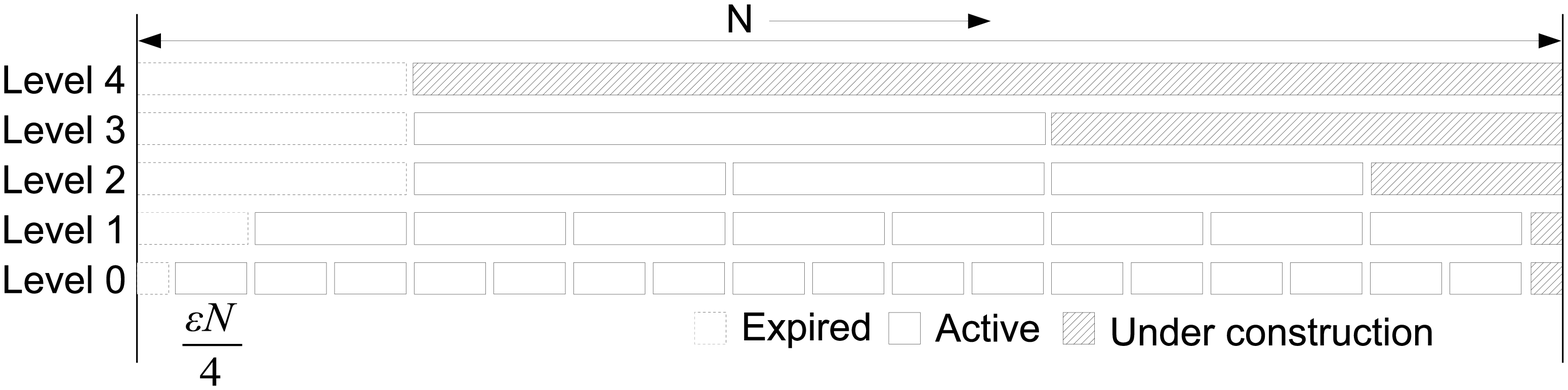}
    \caption{Levels and blocks in the fixed-size sliding window model.}
    \label{fig:FSSW}
\end{figure}

\begin{figure}[htbp]
    \includegraphics[width=\columnwidth]{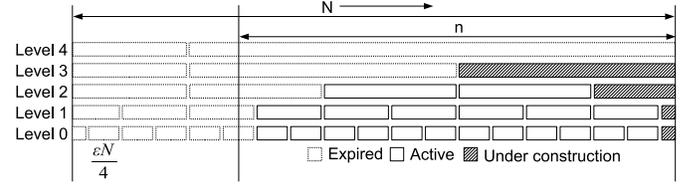}
    \caption{$F_n(N, \epsilon)$, a restriction of $F(N, \epsilon)$ to the last $n$ elements.}
    \label{fig:VSSW}
\end{figure}

Arasu \emph{et al.} \cite{DBLP:conf/pods/ArasuM04} generalized SW model and came up with fixed- and variable- size sliding window approximate quantile algorithms. A window is fixed-size when insertion and deletion of elements 
must appear in pairs after initialization. The fixed-size sliding window model defines blocks and levels as Figure~\ref{fig:FSSW}. 
Each level preserves a partitioning of the data stream into non-overlapping blocks of equal size. Blocks and levels are both numbered 
sequentially. Assuming the window size is $N$, the block $b$ in level $l$ contains a $Summary$ \cite{DBLP:conf/sigmod/GreenwaldK01}, 
denoted as $F(N, \epsilon)$ in this model, which covers elements with arriving positions in the range $[b2^l\frac{\epsilon{N}}{4}, 
(b+1)2^l\frac{\epsilon{N}}{4}-1]$. Similar with SW model, a bucket is assigned one of 3 states at any point of time:
\begin{itemize}
  \item A block is active if all elements covered by it belong to the current window.
  \item A bucket is expired if it covers at least one element which is older than the other $N$ elements.
  \item A bucket is under construction if some of its elements belong to the current window while others are yet to arrive.
\end{itemize}
The highest level with active blocks or blocks under construction $L$ is $\log_2(\frac{4}{\epsilon})$. Blocks in levels above are marked expired.
For each active block in level $l$, a $F(N, \epsilon_l)$ is retained, where $\epsilon_l=\frac{\epsilon}{2(2L+2)}2^{(L-l)}$. 
When a block is under construction, GK01 computes its $F(N, \frac{\epsilon_l}{2})$, and it is converted to a $F(N, \epsilon_l)$ 
in the same way as COMPRESS in SW model. The $F(N, \epsilon_l)$ occupies $O(\frac{1}{\epsilon_l})$ space. Arasu \emph{et al.} proved that using a set 
of $Summaries$ covering $N_1, N_2, ..., N_s$ elements each with approximate error $\epsilon_1,\epsilon_2,...,\epsilon_s$, a 
$\epsilon$-approximate quantile can be computed, where $\epsilon = \frac{\epsilon_1N_1+\epsilon_2N_2+\cdots+\epsilon_sN_s}{N_1+N_2+\cdots+N_s}$. 
Thus, the fixed-size sliding window model can computes $\epsilon$-approximate quantiles over the last $N$ elements using 
$O(\frac{1}{\epsilon}\log\frac{1}{\epsilon}\log{N})$ space.
Contrary to a fixed-size window, a variable-size window bears no limitation on insertion and deletion, so the 
window size keeps changing. Arasu \emph{et al.} used $V(n, \epsilon)$ to denote a $epsilon$-approximate $Summary$ covering $n$ elements, where $n$ is 
the current size of the window. When a new element arrive, it becomes $V(n+1, \epsilon)$, and when the oldest element leaves, it gets 
$V(n-1, \epsilon)$. Besides, $F_n(N, \epsilon)$ is defined as a restriction of $F(N, \epsilon)$ to the last $n$ elements as Figure~\ref{fig:VSSW}.
$F_n(N, \epsilon)$ is the same as $F(N, \epsilon)$ except that only blocks whose elements all belong to the most $n$ recent elements, instead of 
$N$ elements, are assumed active. $V(n, \epsilon)$ can be constructed by a set of $F_n(N, \epsilon)$ in the form of 
$\{F_n(2^k,\frac{\epsilon}{2}),F_n(2^{k-1},\frac{\epsilon}{2}),...,F_n(\frac{2}{\epsilon},\frac{\epsilon}{2})\}$, where $k$ is an integer satisfying 
$2^{k-1}<n\leq 2^k$. $V(n, \epsilon)$ is maintained under various operations as follows:
\begin{itemize}
  \item \textbf{INSERT} Update all $F_n(N, \epsilon)$ in $V(n, \epsilon)$ by incrementing $n$. If $n+1=2^k+1$, create 
  $F_{2^k}(2^{k+1}, \frac{\epsilon}{2})$ from $F_{2^k}(2^k, \frac{\epsilon}{2})$ and insert the new element into it, thus 
  getting $F_{2^k+1}(2^{k+1}, \frac{\epsilon}{2})$.
  \item \textbf{DELETE} Compute $F_{n-1}(2^k, \frac{\epsilon}{2})$ from $F_n(2^k, \frac{\epsilon}{2})$. If $n-1=2^{k-1}$, 
  remove $F_{2^{k-1}}(2^k, \frac{\epsilon}{2})$.
  \item \textbf{QUERY} In order to query $\epsilon$-approximate quantiles over the most $n'\leq n$ recent elements, find the integer $l$ 
  satisfying $2^{l-1}<n\leq 2^l$. Then query $F_n(2^l,\frac{\epsilon}{2})$ and return the answer.
\end{itemize}
The space complexity of the variable-size sliding window model is $O(\frac{1}{\epsilon}\log\frac{1}{\epsilon}\log(\epsilon{N})\log{N})$.

For uncertain data, 
one may still expect to obtain some consistent answers for queries \cite{DBLP:conf/sigmod/LianCS10}.  
Liang \emph{et al.} \cite{DBLP:journals/access/LiangLL19} extended the problem and resolved approximate quantile queries over uncertain data streams. More 
specifically, it focuses on maintaining quantile $Summaries$ \cite{DBLP:conf/sigmod/GreenwaldK01} over data streams whose elements are drawn from individually 
domain space, represented by continuous or discrete pdf. An uncertain data stream $S^u$ is a sequence of elements as $\{e_1^u,e_2^u,...\}$, each of 
which is drawn from a domain $D_i$ with pdf $f_i:D_i\rightarrow{(0,1]}$ such that $\sum_{p\in{D_i}}f_i(p)=1$. So, $S^u$ is a concise representation 
of exponential or infinite number of possible worlds $\mathbb{W}$. Each world $W=\{p_i|pi\in{D_i}, i=1,2,...\}$ is a deterministic stream with 
probability $Pr(W)=\prod_{p_i\in{W}}f_i(p_i)$. Therefore, the definition of approximate quantiles is generalized as the element $p\in{D_i}$ of 
$e_i^u\in{S^u}$ such that $\sum_{W\in\mathbb{W}}Pr(W)q(r,r_W^p)-\sum_{W\in\mathbb{W}}Pr(W)q(r,r_W^{p_{min}})\leq\epsilon{N}$, where $r_W^p$ is the 
rank of $p$ in $W$ and $p_{min}$ is the element minimizing $\sum_{W\in\mathbb{W}}Pr(W)q(r,r_W^p)$. Liang \emph{et al.} proposed 2 error metric 
functions as $q(r,r_W^p)$, the squared error function $q(r,r_W^p)=(r-r_W^p)^2$ and the related error function $q(r,r_W^p)=r-r_W^p$. Following GK01, 
an online algorithm, namely UN-GK, is introduced. UN-GK adjusts tuples in $Summaries$ as $v_i=p_i\in{D_j}$ of $e_j^u\in{S^u}$, 
$g_i=PC_{min}(p_i)-PC_{min}(p_{i-1})$, and $\Delta_i=PC_{max}(p_i)-PC_{min}(p_i)$, where $PC_{min}(p)$ and $PC_{max}(p)$ is the lower bound and the 
upper bound of probabilistic cardinality \cite{DBLP:journals/ijsysc/LiangZ0H15} of elements in $S^u$ no larger than $p$. And (\ref{eq:2-2}) is modified as 
$g_i + \Delta_i\leq2\epsilon{PC(S^u)}$, where $PC(S^u)$ is the probabilistic cardinality of all elements in $S^u$ as $PC(S^u)=|S^u|$. In this way, 
a $\epsilon$-approximate quantile can be queried anytime over an uncertain data stream, and its space complexity is 
$O(\frac{1}{\epsilon}\log(\epsilon{PC(S^u)}))$.

\subsection{Distributed Model}
In distributed models such as sensor networks, communication between nodes consumes much energy and cuts down the battery life of 
power-constrained devices. In addition, data transmission takes most of the running time of algorithms. Therefore, the priority of approximate 
quantile algorithms over distributed models is to reduce the communication cost by decreasing the size of transmitted data.

\begin{figure}[htbp]
    \includegraphics[width=.7\columnwidth]{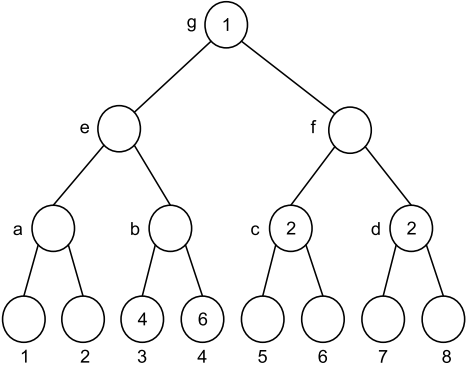}
    \caption{Structure of a q-digest tree, where $n = 15, k = 5, \sigma = 8$.}
    \label{fig:q-digest}
\end{figure}

In 2004, Shrivastava \emph{et al.} \cite{DBLP:conf/sensys/ShrivastavaBAS04} designed an approximate quantile algorithm on distributed sensor networks with 
fixed-universe data, named as q-digest. Fixed-universe data refers to elements from a definite collection. Q-digest uses a unique binary tree 
structure to compress and store elements so that the storage space is cut down. The size of the fixed-universe collection is denoted as $\sigma$ 
and the compress coefficient is denoted as $k$. Figure~\ref{fig:q-digest} is the structure of a q-digest tree, which exists in each network node. 
Each node of the tree covers elements ranging from $v.min$ to $v.max$, recording the sum of their frequencies as $count(v)$. Each leaf 
represents one element in the universe, in other words, $v.min = v.max$. Take node $d$ as an example, its coverage is $[7,8]$ and there are 2 
elements in this range. Besides, each node $v$ of the tree must satisfy:
\begin{equation}
  count(v) \leq \lfloor\frac{n}{k}\rfloor \label{eq:2-6}
\end{equation}
\begin{equation}
  count(v) + count(v_p) + count(v_s) > \lfloor\frac{n}{k}\rfloor \label{eq:2-7}
\end{equation}
where $v_p$ is $v$'s parent node and $v_s$ is its sibling node. (\ref{eq:2-6}) guarantees the upper bound of $v$' coverage to narrow 
approximation error while (\ref{eq:2-7}) demonstrates when to merge two small nodes so that the storage cost can be reduced.

Q-digest proposed an algorithm COMPRESS to merge nodes of a tree:
\begin{enumerate}
  \item Scan nodes from bottom to top, from left to right, and sum up $count(v)$ and $count(v_s)$ when (\ref{eq:2-7}) is violated.
  \item Store the sum in $v_p$. 
  \item Remove the count in $v$ and $v_s$.
\end{enumerate}
In order to reduce data communication, we number nodes in the tree from top to bottom, from left to right and only transmits nonempty 
nodes in the form of $(id(v),count(v))$, in which way every transmission contains only $O(\log{\sigma}+\log{N})$ data. For example, node $c$ 
in Figure~\ref{fig:q-digest} is transformed to $(6, 2)$. When a network node receives q-digest trees from other nodes, it merges them 
with its own tree by adding up $count$ of nodes representing the same elements and applying COMPRESS. After merging all q-digest trees 
in the network, we can query quantiles by traversing the ultimate tree in postorder, summing up $count$ of passed nodes until the sum exceeds 
$\lfloor\phi{N}\rfloor$. The queried quantile is $v.max$ of the current node. The total communication cost of q-digest is 
$O(\frac{1}{\epsilon}|v|\log\sigma)$, where $|v|$ denotes the number of network nodes.

In the same year, Greenwald \emph{et al.} \cite{DBLP:conf/pods/GreenwaldK04} proposed an algorithm on distributed models, referred to as GK04. Unlike q-digest, 
GK04 does not require that all data be fixed-universe. It maintains a collection of $\frac{\epsilon}{2}$-approximate 
$Summaries$ \cite{DBLP:conf/sigmod/GreenwaldK01}, denoted as $S_v$, in each node $v$ in the 
sensor network. More specifically, $S_v=\{S^1_v, S^2_v,..., S^k_v\}$, where $S^i_v$ covers $n^i_v$ elements, and $S^i_v$ is classified 
as $class(S^i_v)=\lfloor\log{n^i_v}\rfloor$. In the network, data are transmitted in the form of $S_v$. When data arrives at a node, 
it combines its own $S_v$ with the coming $S_v$ by iteratively merging $S_v$ of the same class from bottom to top. Once the iteration ends, 
a pruning algorithm is applied to $S^i_v$ to reduce the number of its tuples to $\log\frac{n_v}{\epsilon} + 1$ at most. The pruning would bring 
up the approximation error of $S^i_v$ from $\epsilon'$ to $\epsilon' + \frac{\epsilon}{2\log{n_v}}$ at most, so a $\epsilon$-approximate $S$ is 
computed after merging $S_v$ from all nodes. GK04 only transmits $O(\frac{1}{\epsilon}\log^2{N})$ data, where $N$ is the number of all data 
in the network. Moreover, if the height of the network topology is far smaller than $N$, Greenwald \emph{et al.} improved GK04 to reduce the total 
communication cost to $O(\frac{|v|}{\epsilon}\log{N}\log(\frac{h}{\epsilon}))$. Its basic idea is to apply a new operation, REDUCE, which 
makes sure that all nodes transmits less than $O(\log\frac{h}{\epsilon})$ $Summaries$, after merging and pruning. 

Q-digest and GK04 are both offline algorithms that compute quantiles over stationary data in distributed 
models. Besides, there are also algorithms focusing on distributed models whose nodes receiving continuous 
data streams. In 2005, Cormode \emph{et al.} \cite{DBLP:conf/sigmod/CormodeGMR05} proposed a quantile algorithm for the scenario 
that multiple mutually isolated sensors are connected with one coordinator, which traces updating quantiles 
in real time. Its goal is to ensure $\epsilon$-approximate quantiles at the coordinator while minimizing 
communication cost between nodes and the coordinator. In general, each remote node maintains a local approximate 
$Summary$ \cite{DBLP:conf/sigmod/GreenwaldK01} and informs the coordinator after certain number of updates. In the coordinator, 
the approximation error $\epsilon$ is divided into 2 parts as $\epsilon=\alpha+\beta$:
\begin{itemize}
  \item $\alpha$ is the approximation error of local $Summaries$ sent to the coordinator.
  \item $\beta$ is the upper bound on the deviation of local $Summaries$ since the last communication.
\end{itemize}
Intuitively, larger $\beta$ allows for large deviation, thus less communication between nodes and the coordinator. 
But because $\epsilon$ is fixed, $\alpha$ is smaller, increasing the size of $Summaries$ sent to the coordinator 
each time. In other words, $\alpha$ and $\beta$ are on the trade-off relationship. To resolve the trade-off, 
Cormode \emph{et al.} introduces a prediction model in each remote node that captures the anticipated behavior of its local 
data stream. With the model, the coordinator is able to predict the current state of a local data stream while 
computing the global $Summary$ and the remote node can check for the deviation between its $Summary$ and 
the coordinator's prediction. The algorithm proposes 3 concise prediction models:
\begin{itemize}
  \item \textbf{Zero-Information Model} assumes that there is no local update at any remote node since the last communication.
  \item \textbf{Synchronous-Updates Model} assumes that at each time step, each local node receives one update to its distribution. 
  \item \textbf{Update-Rates Model} assumes that updates are observed at each local node at a uniform rate with a notion of global time. 
\end{itemize}
Using prediction models to reduce communication times, the total communication cost of this algorithm is 
$O(\frac{|v|}{\epsilon^2}\log{N})$. 

In 2013, Yi \emph{et al.} \cite{DBLP:journals/algorithmica/YiZ13} proposed an algorithm in the same scenario and optimized the cost to 
$O(\frac{|v|}{\epsilon}\log{N})$. The algorithm divides the whole tracking period into $O(\log{N})$ rounds. A 
new round begins whenever $N$ doubles. It first resolves the median-tracking problem, which can be easily generalized 
to the quantile-tracking problem. Assume that $M$ is the cardinality of data, fixed at the beginning of a round, and 
$m$ is the tracking median at the coordinator. The coordinator maintains 3 data structures. The first one is 
a dynamic set of disjoint intervals, each of which contains between $\frac{\epsilon{M}}{8}$ and $\frac{\epsilon{M}}{2}$ elements. 
The others are 2 counters $C.\Delta(L)$ and $C.\Delta(R)$, recording the number of elements received by all 
remote nodes to the left and the right of $m$ since last update respectively. They are guaranteed with an 
absolute error at most $\frac{\epsilon{M}}{8}$ by asking each remote node to send an update whenever it receives 
$\frac{\epsilon{M}}{8|v|}$ elements to the left or the right of $m$. When 
$|C.\Delta(L) - C.\Delta(R)| \geq \frac{\epsilon{M}}{2}$, $m$ is updated as follows:
\begin{enumerate}
  \item Compute the total number of elements to the left and the right of $m$, $C.L$ and $C.R$ and let $d = \frac{1}{2}|C.L-C.R|$.
  \item Compute the new median $m'$ satisfying that $|r(m) - r(m') - d| \leq \frac{\epsilon{M}}{4}$, where $r$ is the rank of element in all data. Replace $m$ with $m'$.
  $m'$ can be found quickly with the set of intervals. First find the first separating element $e_1$ of the intervals to the left of $M$. 
  Then compute $n_1$, the number of elements at all remote sites that are in the interval $[e_1, m]$. 
  If $|n_1-d|\leq\frac{\epsilon{M}}{2}$, $e_1$ is $m'$. Otherwise find the next separating element $e_i$ and 
  count the number $n_i$ until $|n_i-d|\leq\frac{\epsilon{M}}{2}$, then set $m'$ to $e_i$.
  \item Set $C.\Delta(L)$ and $C.\Delta(R)$ to $0$.
\end{enumerate}
Yi \emph{et al.} has proved that $m$ is at most $\epsilon{M}$ elements away from the ground truth. Step 1 needs to exchange 
$O(|v|)$ messages. As for Step 2, $m'$ can be found after at most $O(1)$ searches and the cost of each search is 
$O(|v|)$, making the total cost $O(|v|)$. Because each update increases $N$ by a factor of $1+\frac{\epsilon}{2}$, $m$ 
is updated at most $O(\frac{1}{\epsilon})$ times. So, the total communication cost is $O(\frac{|v|}{\epsilon})$ each 
round and $O(\frac{|v|}{\epsilon}\log{N})$ for the whole algorithm.

\section{Randomized Algorithms}
\label{sect:randomized algorithms}
Generally, randomized approximate quantile algorithms are combined with random sampling. They first sample a part of data and compute overall 
approximate quantiles with this portion. With less data taken into computation, the computation cost is cut down. In fact, many deterministic 
quantile algorithms propose randomized versions in this way \cite{DBLP:conf/sigmod/MankuRL99, DBLP:conf/pods/ArasuM04, DBLP:journals/algorithmica/YiZ13}.

\subsection{Streaming Model}
Back to 1971, Vapnik \emph{et al.} \cite{vapnik2015uniform} proposed a randomized quantile algorithm with space complexity of $O(\frac{1}{\epsilon^2}\log\frac{1}{\epsilon})$. 
This benchmark were raised to $O(\frac{1}{\epsilon}\log^2\frac{1}{\epsilon})$ by Manku \emph{et al.} \cite{DBLP:conf/sigmod/MankuRL99} in 1999, referred to as MRL99. 
MRL99 does not require prior knowledge of the data size $N$ and occupies less space in experiments when $\phi$ is an extreme value. The algorithm 
are improved based on MRL98 \cite{DBLP:conf/sigmod/RajagopalanML98} so they have the identical frame with only minor differences in the operation NEW: 
\begin{enumerate}
  \item For each buffer, randomly select an element among $r$ consecutive elements.
  \item Repeats this operation $k$ times to get $k$ initial elements. 
  \item Set the buffer's weight to $r$.
\end{enumerate}
Notice that MRL99 equals with MRL98 if $r = 1$. Because of the collapsing strategy as Figure~\ref{fig:MRL98-schema}, 
the larger weight a buffer has, the greater chance there its data must be retained while collapsing. If $r$ is kept consistent, 
the probability of newly arrived data being selected will go down continuously. So $r$ should keep changed dynamically, meaning the sampling 
is nonuniform. MRL99 initially sets $r$ to 2 and traces a parameter $h$, representing the maximum height of all buffers. When a buffer's 
height reaches $h + i$ for the first time, where $i \geq 0$, $r$ is doubled. By tuning $h$, $b$ and $k$, MRL99 manages to compute approximate 
quantiles with a probability of at least $1 - \delta$.

Recalling two sliding window models in Section~\ref{sect:deterministic-streaming}, Arasu \emph{et al.} proposed the fact 
that the quantile of a random sample of size $O(\frac{1}{\epsilon^2}\log\delta^{-1})$ is an $\epsilon$-approximate 
quantile of $N$ elements with the probability at least $1-\delta$. To sample elements of specific size, a fast 
alternative is to randomly select one out of $2^k$ successive elements, where $k = \lfloor\log_2N/(\frac{1}{\epsilon^2}\log\delta^{-1})\rfloor$. 
In this way, $k$ grows logarithmically along with the data stream as required, so the approximation error and the randomization degree 
are guaranteed.

Another algorithm was proposed by Agarwal \emph{et al.} \cite{DBLP:journals/tods/AgarwalCHPWY13}, which is based on $Summaries$ \cite{DBLP:conf/sigmod/GreenwaldK01}. 
Its basic idea is to sample tuples from multiple $Summaries$ and merge them to compute approximate quantiles with 
low space complexity. There are two situations while merging: same-weight merges and uneven-weight merges. Same-weight merges are for merging two 
$S$s covering the same number of elements. It contains following steps: 
\begin{enumerate}
  \item Combine the two $S$s in a sorted way.
  \item Label the tuples in order and classify them by label parity.
  \item Equiprobably select one class of tuples as the merged result $S_{merged}$.
\end{enumerate}
Assuming each $S$ covers $k$ elements, if we have $k = O(\frac{1}{\epsilon}\sqrt{\log(\frac{1}{\epsilon\delta})})$, the algorithm will answer 
quantile queries with a probability of at least $1 - \delta$. Uneven-weight merges are for merging two $S$s of different sizes, which can be 
reduced to same-weight merges by a so-called logarithmic technique \cite{DBLP:conf/pods/GreenwaldK04}. The space complexity of the algorithm 
is $O(\frac{1}{\epsilon}\log^{1.5}\frac{1}{\epsilon})$.

However, Agarwal \emph{et al.} just proposed and analyzed the algorithm in theory without implementation. Afterwards, Felber \emph{et al.} \cite{DBLP:conf/approx/FelberO15} 
came up with a randomized algorithm whose space complexity is $O(\frac{1}{\epsilon}\log\frac{1}{\epsilon})$ but also did not realize it. 
Besides, this algorithm is not actually useful but only suitable for theoretical study because its hidden coefficient of $O$ is too large.

In 2016, Karnin \emph{et al.} \cite{DBLP:conf/focs/KarninLL16} achieved the space complexity of $O(\frac{1}{\epsilon}\log\log\frac{1}{\epsilon\delta})$. 
Its idea is based on MRL99 with some improvements. 
The first improvement is using increasing compactor capacities as the height gets larger (a compactor operates a COLLAPSE process in MRL99). 
The second comes from special handling of the top $\log\log\frac{1}{\delta}$ compactors.
And the last is replacing those top compactors with $Summaries$ \cite{DBLP:conf/sigmod/GreenwaldK01}.
 
\subsection{Distributed Model}
As for distributed models, Huang \emph{et al.} \cite{DBLP:conf/sigmod/HuangWYL11} proposed a randomized quantile algorithm which brings down total 
communication cost from $O(|v|\log^2\frac{N}{\epsilon})$ in GK04 \cite{DBLP:conf/pods/GreenwaldK04} to $O(\frac{1}{\epsilon}\sqrt{|v|h})$, where $h$ 
denotes the height of the network topology. It contains two version: the flat model and the tree model. For the flat model, all other nodes 
are assumed to be directly connected to the root node. The algorithm is designed as following steps: 
\begin{enumerate}
  \item Sample elements in node $v$ with a probability of $p$ and compute their ranks in $v$, denoted as $r(a,v)$ where $a$ is a sampled element.
  \item Transmit sampled elements, as well as ones from its child nodes, to its parent node $v_p$.
  \item Find predecessors of $a$, denoted as $pred(a, v_s)$, in $v$'s sibling nodes $v_s$.
  \item Estimate the rank of $a$ in $v_s$, denoted as $\hat{r}(a,v_s)$, according to $r(pred(a, v_s), v_s)$.
  \item Compute the approximate rank of $a$ in $v_p$ as $r(a, v_p) = \sum\hat{r}(a, v_s) + r(a, v)$.
\end{enumerate}
If elements are not uniformly distributed in the network and some nodes contain the majority, the total communication cost will increase dramatically 
as the effect of load imbalance. The algorithm resolves the problem by tuning $p$ based on data amount in each node:
\begin{itemize}
  \item If the amount is greater than $N/\sqrt{|v|}$, $p$ is set to $1/(\epsilon{N_v})$.
  \item Otherwise, $p$ is set to $\Theta(\sqrt{|v|}/(\epsilon{N}))$.
\end{itemize}
The inconsistent probability of sampling makes sure that $O(\frac{1}{\epsilon})$ elements are sampled at most in each node no matter how many elements there 
exist at first. After transmitting all sampled elements to the root node, the element $a$ whose rank $r(a, v_{root})$ is closest to 
$\lfloor\phi{N}\rfloor$ is returned as the queried quantile. For the tree model, things become more complicated for two reasons. First, an intermediate 
node may suffer from heavy traffic going through if it has too many descendants without any data reduction. Second, each message needs $O(h)$ hops to reach the root node, leading to 
the total communication of $O(h\sqrt{|v|}/\epsilon)$. To resolve the first problem, Huang \emph{et al.} proposed a algorithm MERGE in a systematic way to reduce 
data size. As for the second problem, the basic idea is to partition the routing tree into $t$ connected components, each of which has $O(|v|/t)$ nodes. 
Then each component is shrunk into a "super node". Now the height of the tree reduces to $t$. By setting $t = |v|/h$, the desired space bound becomes 
$O(\frac{1}{\epsilon}\sqrt{|v|h})$.

In 2018, Haeupler \emph{et al.} \cite{DBLP:conf/podc/HaeuplerMS18}, gave a drastically faster gossip algorithm, referred as 
Haeupler18, to compute approximate quantiles. Gossip algorithms \cite{DBLP:conf/infocom/BoydGPS05} are algorithms that allow nodes 
in a distributed network to contact with each other randomly in each round and gradually converge to get final results. The algorithm 
contains two phases. In the first phase, each node adjusts its value so that the quantiles around $\phi$-quantile become the median 
quantiles approximately. And in the second phase, nodes compute their approximate median quantiles to get the global result. 
Haeupler \emph{et al.} proved that the algorithm requires $O(\log\log{N}+\log\frac{1}{\epsilon})$ rounds to solve the $\epsilon$-approximate 
$\phi$-quantile problem with high probability.

\section{Improvement}
\label{sect:improvement}
So far, in the discussion about approximate quantile algorithms, they are generally used with constant 
approximation error and indiscriminate performance on data regardless of data distribution. However, in some 
cases, we may have known that the data is skewed \cite{DBLP:conf/icde/CormodeKMS05, DBLP:conf/pods/CormodeKMS06, DBLP:conf/pimrc/LinLL17, DBLP:journals/access/LiuZLL18, DBLP:journals/corr/abs-1902-04023}. In other cases, quantile queries over high-speed data streams need to be updated and answered highly efficiently \cite{DBLP:journals/tkde/LinXZLYZY06, DBLP:conf/ssdbm/ZhangW07}. In addition, there are also techniques for optimizing quantile computation with the help of GPUs \cite{DBLP:conf/sigmod/GovindarajuRM05, zhou2010parallel}. This section presents several techniques for improving the performance and efficiency of approximate quantile algorithms in various scenarios.

\subsection{Skewness}
The first algorithm is known as t-digest, proposed by Dunning \emph{et al.} \cite{DBLP:journals/corr/abs-1902-04023}. T-digest is for computing extreme quantiles 
such as the 99th, 99.9th and 99.99th percentiles. Totally different from q-digest, its basic idea is to cluster real-valued samples like 
histograms. But they differ in three aspects. First, the range covered by clusters may overlap. Second, instead of lower and upper bounds, a 
cluster is represented by a centroid value and an accumulated size on behalf of the number of elements. Third, clusters whose range is close 
to extreme values contain only a few elements so that the approximation error is not absolutely bounded, but relatively bounded, which is 
$\phi(1 - \phi)$. T-digest can be applied to both streaming models and distributed models because the proposed cluster is a mergeable 
structure. The merge is restricted by the size bound of clusters. Dunning \emph{et al.} proposed several scale functions to define the bound. 
The standard is that the size of each cluster should be small enough to get accurate quantiles, but large enough to avoid winding up too many 
clusters. A scale function is 
\begin{equation}
  f(\phi) = \frac{\delta}{2\pi}\sin^{-1}(2\phi-1) \label{eq:4-1}
\end{equation}
where $\delta$ is the compression parameters and the size bound is defined as 
\begin{equation}
  W_{bound} = f(\frac{W_{left} + W}{N}) - f(\frac{W_{left}}{N}) \leq 1 \label{eq:4-2}
\end{equation}
where $W_{left}$ and $W$ are respectively the weight of clusters whose centroid values are smaller than that of the current cluster and of 
current cluster.
\begin{figure}[htbp]
    \centering
    \includegraphics[width=0.55\columnwidth]{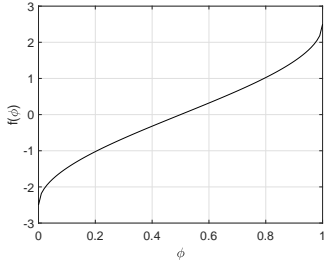}
    \caption{Function of (\ref{eq:4-1}) with $\delta = 10$.}
    \label{fig:function-asin}
\end{figure}
As Figure~\ref{fig:function-asin} shows, (\ref{eq:4-1}) is non-decreasing and is steeper when $\phi$ is closer to $0$ or $1$, which means 
clusters covering extreme values have smaller size, making the algorithm more accurate for computing extreme quantiles and more robust 
for skewed data.

The second algorithm, proposed by Lin \emph{et al.} \cite{DBLP:conf/pimrc/LinLL17} and elaborated by Liu \emph{et al.} \cite{DBLP:journals/access/LiuZLL18}, 
is aimed at streaming models, using nonlinear interpolation. The algorithm maintains two buffers, the quantile buffer $Q=\{q_1,q_2,...,q_m\}$, where $q_i$ 
is the approximate $\phi_i$-quantile, and the data buffer $B$ of size $n$, holding the most recent $n$ elements. $Q$ is estimated from observed data and 
incrementally updated when $B$ is full-filled. In order to estimate the extreme quantiles accurately, the nonlinear interpolation $F(x)$, which is an 
approximate distribution function estimated from a training set stream, is leveraged together with $B$ and $Q$ to update $Q$.

The third algorithm was proposed by Cormode \emph{et al.} \cite{DBLP:conf/icde/CormodeKMS05} for skewed data. 
Again, the algorithm is an improved version of GK01 for two problems. 
The first problem is the biased quantiles problem. 
Low-biased quantiles are the set of elements whose rank $\lfloor\phi^iN\rfloor$ for $i = 1,2,...,\log_{1/\phi}N$, 
and high-biased quantiles are symmetry by reversing the ordering relation. 
The definition is easy to be generalized to approximate quantiles. The problem is computing the first $k$ elements in high-biased 
approximate quantiles. Similar with GK01 that $g_i$ and $\Delta_i$ are restricted by $g_i + \Delta_i \leq \lfloor 2\epsilon{N} \rfloor$, 
the algorithm generalizes the restriction as $g_i + \Delta_i \leq f(r_i, N)$, where $f(r_i, N)$ is an appropriate function. $r_i$ is the 
rank of $v_i$ equaling $\sum_{j=1}^{i-1}g_j$. For the biased quantiles problem, $f(r_i, N)$ is set to $2\epsilon{r_i}$. The restriction is 
tighter than that in GK01 so the correctness is guaranteed. Cormode \emph{et al.} proved that the space lower bound is 
$\Omega(\frac{1}{\epsilon}min(k\log\frac{1}{\phi}, \log(\epsilon{N})))$. The other problem is targeted quantiles problem that quantiles 
meeting a set of pairs $T = \{(\phi_i, \epsilon_i)\}$ are required to be maintained. In such case, $f(r_i, N)$ is set to 
$\frac{2\epsilon_ir_i}{\phi_i}$ if $\phi_iN \leq r_i \leq N$ and $\frac{2\epsilon_i(N-r_i)}{1-\phi_i}$ if $0 \leq r_i \leq \phi_iN$. 
Also, the state-of-the-art space upper bound for biased quantile computation is $O(\frac{1}{\epsilon}\log^3{\epsilon{N}})$ with a deterministic
comparison-based merging-and-pruning strategy \cite{DBLP:conf/cikm/ZhangW07}. 
As for a randomized version, Zhang \emph{et al.} \cite{DBLP:conf/icde/ZhangLXKW06} achieved expected space of 
$O(\frac{1}{\epsilon}\log(\frac{1}{\epsilon}\log\frac{1}{\epsilon})\frac{\log^{2+\alpha}{\epsilon{N}}}{1-2^{-\alpha}})$ 
where $\alpha>0$ and worst case of $O(\frac{1}{\epsilon^2}\log\frac{1}{\epsilon}\log^2{\epsilon{N}})$. 

For problems resolved by q-digest \cite{DBLP:conf/sensys/ShrivastavaBAS04} that all elements are selected from a fixed-universe collection, 
Cormode \emph{et al.} \cite{DBLP:conf/pods/CormodeKMS06} combined the binary tree structure in q-digest and standard dictionary data structures \cite{DBLP:books/mg/CormenLR89}, 
proposing a new deterministic algorithm to compute biased quantiles with space complexity of $O(\frac{1}{\epsilon}\log{\sigma}\log(\epsilon{N}))$, 
where $\sigma$ denotes the size of the fixed-universe collection. 
And a simpler sampling-based approach by Gupta \emph{et al.} \cite{DBLP:conf/soda/GuptaZ03} uses space of $O(\epsilon^{-3}\log^2N\log\sigma)$.

\subsection{High-Speed Data Streams}
\begin{figure}[htbp]
    \includegraphics[width=\columnwidth]{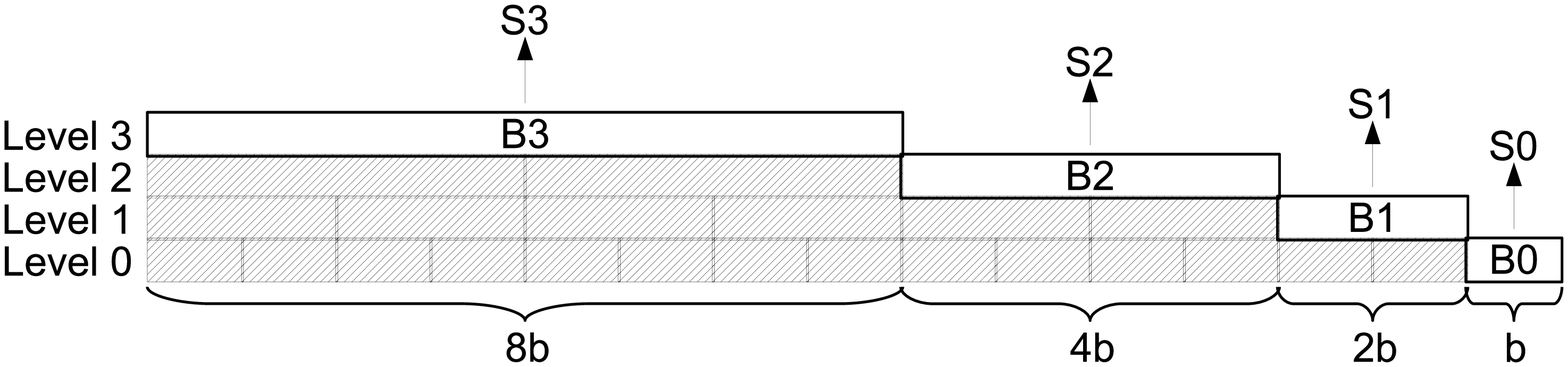}
    \caption{A multi-level summary structure with $L = 3$.}
    \label{fig:high-speed}
\end{figure}

In order to compute quantiles over high-speed data streams, both computational cost and per-element update cost 
need to be low. Zhang \emph{et al.} \cite{DBLP:conf/ssdbm/ZhangW07} proposed an algorithm for both fixed- and arbitrary- size 
high-speed data streams. Let $N$ and $n$ denote the number of elements in the entire data stream and elements 
seen so far. For the fixed-size data streams, where $N$ is given, a multi-level summary structure 
$S=\{s_0,s_1,...,s_L\}$ is maintained, where $s_i$ is the summary at level $i$, as shown in Figure~\ref{fig:high-speed}. 
Each element in $s$ is stored with its upper and lower bound of rank, $r_{min}(e)$ and $r_{max}(e)$.
The data stream is divided into blocks of size $b=\lfloor\frac{1}{\epsilon}\log\epsilon{N}\rfloor$ and $s_i$ covers 
a disjoint bag $B_i$. Among bags, $B_0$ contains the most recent blocks even though it may be incomplete. The structure 
is maintained as follows:
\begin{enumerate}
  \item Insert the new element to $s_0$.
  \item If $|s_0| < b$, the procedure is done. Otherwise, compress $s_0$ to generate a sketch $s_c$ 
  of size $\lfloor\frac{b}{2}\rfloor$ and send it to level $1$. COMPRESS would raise the 
  approximation error from $\epsilon_0$ to $\epsilon_0+\frac{1}{b}$.
  \item If $s_1$ is empty, set $s_1$ to $s_c$ and the procedure is done. Otherwise, merge $s_1$ with $s_c$ and 
  empty $s_0$. Finally, compress the merged $s$ and send it to level $2$.
  \item Repeat the steps above until an empty $s_i$ is found.
\end{enumerate}
To answer quantiles, the algorithm first sorts $s_0$ and merges $s$ at all levels. Then it searches the merged $s$ 
and return the element satisfying that $r_{min}(e) \geq r-\lfloor\epsilon{N}\rfloor$ and $r_{max}(e) \leq r+\lfloor\epsilon{N}\rfloor$.
As for the arbitrary-size data streams, the basic idea is to partition the data stream into disjoint sub-streams $d_i$ 
with the size $\frac{2^i}{\epsilon}$ in their arrival order, and use the algorithm for the fixed-size data streams on each sub-stream 
because their length is known now. The computational cost and the per-element update cost of both algorithms are 
$O(N\log(\frac{1}{\epsilon}\log\epsilon{N}))$ and $O(\log\log{N})$. Compared to GK01, the experimental results over 
high-speed data streams are reported to achieve about $200\sim300x$ speedup.

Besides, in order to lighten the burden of massive continuous quantile queries with different $\phi$ and $\epsilon$, 
Lin \emph{et al.} \cite{DBLP:journals/tkde/LinXZLYZY06} proposed 2 techniques for processing queries. The first technique is to 
cluster multiple queries as a single query virtually while guaranteeing accuracy. Its basic idea is to cluster 
the queries that share some common results. The second technique is to minimize both the total number of times 
for reprocessing and the number of clusters. It adopts a trigger-based lazy update paradigm.

\subsection{GPU}
Govindaraju \emph{et al.} \cite{DBLP:conf/sigmod/GovindarajuRM05} studied optimizing quantile computation using graphics processors, 
or GPU for short. GPUs are well designed for rendering and allow many rendering applications to raise memory performance \cite{rubinstein2005gpu}. In order to utilize the high computational power of GPUs, Govindaraju \emph{et al.} proposed an algorithm based on sorting networks. Sorting networks are a set of sorting algorithms mapped well to mesh-based architectures \cite{DBLP:conf/afips/Batcher68}. Operations in the algorithm, including comparisons and comparator mapping, are realized by color blending and texture mapping in GPUs. The theoretical algorithm they used is GK04 \cite{DBLP:conf/pods/GreenwaldK04} and by taking advantage of high computational power and memory bandwidth of GPUs, the algorithm offers great performance for quantile computation over streaming models.

\section{Approximate Quantile Computation Tools}
\label{sect:tools}
Over the past decades, many off-the-shelf, open-source packages or tools for quantile computation have been developed and 
available to users. Some of them are based on exact quantile computation while others implement approximate quantile 
computation. In this section, we review such tools, focusing on their algorithmic theories and application scenarios.

Most basically, SQL provides a window function $ntile$\footnote{https://www.sqltutorial.org/sql-window-functions/sql-ntile/} 
which helps with quantile computation. It receives a parameter $b$ and partitions a set of data into $b$ buckets equally. 
Therefore, with $sortBy$ and $ntile$, we can partition the data in order and get the leading element of each partition as the quantile. 
Also, these quantiles are exact.

MATLAB is a famous numerical computing environment and programming language. It provides comprehensive packages for computing 
various statistics or functions. Quantiles are included as its in-tool function\footnote{https://www.mathworks.com/help/stats/quantile.html}. 
The function $quantile$ receives a few parameters, including a numerical array, the cumulative probability and the computation 
kind. It computes quantiles in either exact or approximate way. It computes exact quantiles by the classic algorithm that uses 
sort. And it implements t-digest \cite{DBLP:journals/corr/abs-1902-04023} for approximate quantiles. So, this function is suitable to compute quantiles 
over distributed models, which benefit from parallel computation. 

The distributed cluster-computing framework, Spark, provides approximate quantile computation since version 
2.0\footnote{https://spark.apache.org/docs/2.0.0}. It implements GK01 and can be called in 
many programming languages such as Python, Scala, R and Java. Computing on a Spark dataframe, the function contains 
3 parameters as a dataframe, the name of a numerical column, a list of quantile probabilities and the approximation error. 
Since version 2.2, the function has been upgraded to support computation over multiple dataframe columns.

In Java, Google publishes an open-source set of core libraries, named as Guava\footnote{https://github.com/google/guava}. It includes new collection types, graph libraries, 
support for concurrency, etc. Also, quantile computation is included in its functions as $median()$ and $percentiles()$. 
The quantiles are exact results so the average time complexity of its implementation is $O(n)$ while the worst case 
time complexity is $O(N^2)$. It optimizes multiple quantile computation on the same dataset with indexes, improving 
the performance in some degree. Another package that supports quantile computation is Apache Beam\footnote{https://beam.apache.org/}. It is a unified programming model for 
batch and streaming data processing on execution engines. Unlike Guava, it implements approximate quantile computation.

The programming language, Rust\footnote{https://www.rust-lang.org/}, implements approximate quantile computation over data streams with a moderate amount of 
memory. It implements the algorithms, GK01 \cite{DBLP:conf/sigmod/GreenwaldK01} and CKMS \cite{DBLP:conf/icde/CormodeKMS05}, so that no 
boatload of information is stored in memory. Besides, it implements a variant of quantile computation, an $\epsilon$-approximate frequency 
count for a stream, outputting the $k$ most frequent elements, with the algorithm Misra-Gries \cite{DBLP:journals/scp/MisraG82}, which helps with 
an estimation of quantiles as well. As for C++, a peer-reviewed set of core libraries, Boost, provides exact quantile computation as its in-tool function\footnote{https://www.boost.org/}.

\section{Future Directions and Conclusions}
\label{sect:conclusions}
In this survey, we have presented a comprehensive survey of approximate quantile computation. 
Readers who want to know more about the performance of basic quantile algorithms can read a paper by Luo \emph{et al.} 
\cite{DBLP:journals/vldb/LuoWYC16}, which implements a part of the quantile algorithms above and compares them by experiments. Besides, Cormode \emph{et al.} \cite{DBLP:journals/corr/abs-1905-03838} summarized lower bounds of the researches about comparison-based quantile computation so far and proved that the space complexity of GK01 is optimal by showing a matching lower bound. 

Even though approximate quantiles have been studied for more than three decades, there is still plenty of room for improvement. 
The explosion of data also brings new challenges to this field. There are many studies to resolve quantile computation in 
large-scale data, but not enough. Besides, with the development of industries, new scenarios keep appearing so that quantile computation 
should be optimized specifically as well. And the evolvement of techniques brings new direction and new potential to 
quantile computation. Here we present a few future directions for quantile computation.

First, almost all quantile algorithms require at least one-pass scan over the entire dataset, no matter it is applied on streaming models or 
distributed models. However, at some time, the cost of scanning the entire data is intolerable if quantile queries are demanded to 
be answered in time. In such case, only a portion of the entire dataset is permitted into the process of the whole algorithm. 
This condition is more restricted than that in existing randomized quantile algorithms. Even if randomized algorithms sample a 
part of data to save the space of computation, the process of sampling requires the participation of the entire dataset, 
which means that they need to scan all data once. To resolve the problem, we need to determine the sampling methods first. 
Unlike fine-grained sampling in existing randomized algorithms, we may use coarse-grained sampling, such as sampling in 
the unit of block. The method should also take the way of storage into account. For example, if the dataset is 
stored in distributed HDFS \cite{borthakur2008hdfs}, we may sample part of blocks, avoiding scanning the entire data. 
The sampling methods may be exclusive, varying based on applications. After determining sampling methods, we need to consider 
how much data is sampled to balance the accuracy and the occupation of time and space. One direction is to simulate the 
idea in machine learning \cite{DBLP:journals/jei/BishopN07} that data are trained (or sampled in our situation) continuously until the 
quantile result converges.

Second, many new computation engines are being developed. They can be classified into two categories in general. One is 
streaming computation engines and the other is distributed computation engines. Streaming computation engines, represented by 
Spark Streaming, Storm and Flink \cite{DBLP:conf/ipps/ChintapalliDEFG16}, are aimed at real-time streaming needs with minor 
difference in some ways. For example, Storm and Flink behave like true streaming processing systems with 
lower latencies while Spark Streaming can handle higher throughput at the cost of higher latencies. Distributed computation 
engines, represented by Spark and GraphLab \cite{DBLP:conf/bigdataservice/Wei0ZZH16}, are designed to analyze distributed data such 
as datasets with graph properties.They have their pros and cons in heterogeneous scenarios as well. The characteristics of these engines 
may be utilized in the implementation of algorithms. As reviewed in Section~\ref{sect:tools}, only a small fraction of 
approximate quantile algorithms is implemented in them such as GK01 in Spark. But many other algorithms are still needed to be 
implemented and optimized purposefully. For example, Spark \cite{DBLP:conf/hotcloud/ZahariaCFSS10} is a framework for computation in 
distributed clusters, supporting parallel computation naturally and having potential of benefitting many quantile algorithms such 
as MRL98. Its streaming version, Spark Structured Streaming \cite{DBLP:conf/sigmod/ArmbrustDTYZX0S18}, supports stream processing, as well 
as window operations. It may help quantile algorithms over streaming models, such as Lin SW and Lin n-of-N \cite{DBLP:conf/icde/LinLXY04}, 
to be implemented in a more efficient way, even though the theoretical complexity remains the same. Many quantile algorithms 
are implemented with basic languages while others are even without implementation. Transplanting them to new computation engines 
is not an easy work and there is great room for optimization.

Third, with the appearance of new specific application scenarios, the requirements of approximate quantile algorithms evolve as well. 
For example, nowadays, more and more attention is being paid to the correlation of data, and data graph is one way to present the 
correlation. In a data graph, each edge is usually associated with a weight, representing frequencies of the appearance of the correlation. 
One may need to determine a threshold for pruning edges by their weights for better performance in analysis \cite{DBLP:conf/sigmod/ZhuSL0Z14,DBLP:journals/tkde/GaoSZWLZ18}. 
And we can use quantiles to determine the threshold. However, unlike random sampling in a dataset, the edges in a graph are correlated 
(the frequencies of two edges connecting to the same vertex are correlated). Maybe it is a challenge, as well as an opportunity, 
to efficiently compute approximate quantiles in a correlated data graph. Furthermore, if the graph is constrained by some 
patterns \cite{DBLP:conf/icde/ZhuSWYS14,DBLP:journals/tkde/SongGWZWY17}, the algorithms may be improved and optimized correspondingly. Other cases include computing 
quantiles in the Blockchain network \cite{DBLP:conf/sp/ZyskindNP15}. Unlike traditional distributed models, which have a master node and multiple slave nodes, the 
Blockchain network is decentralized, making merging quantile algorithms infeasible. Further research is needed for computing approximate 
quantiles in such network. Haeupler \emph{et al.} \cite{DBLP:conf/podc/HaeuplerMS18}, which uses gossip algorithms, provides a good direction, 
but there is much more to be done.

%=====================================================

\bibliographystyle{ieeetr}
%\bibliography{main}
%=====================================================

\end{document}